# Point-in-Convex Polygon and Point-in-Convex Polyhedron Algorithms with O(1) Complexity using Space Subdivision

## Vaclav Skala

*Department of Computer Science and Engineering, Faculty of Applied Sciences, University of West Bohemia, Univerzitni 8, CZ 306 14 Plzen, Czech Republic*

**Abstract.** There are many space subdivision and space partitioning techniques used in many algorithms to speed up computations. They mostly rely on orthogonal space subdivision, resp. using hierarchical data structures, e.g. BSP trees, quadtrees, octrees, kd-trees, bounding volume hierarchies etc. However in some applications a non-orthogonal space subdivision can offer new ways for actual speed up. In the case of convex polygon in $E^2$ a simple Point-in-Polygon test is of the $O(N)$ complexity and the optimal algorithm is of $O(\log N)$ computational complexity. In the $E^3$ case, the complexity is $O(N)$ even for the convex polyhedron as no ordering is defined.

New Point-in-Convex Polygon and Point-in-Convex Polyhedron algorithms are presented based on space subdivision in the preprocessing stage resulting to $O(1)$ run-time complexity. The presented approach is simple to implement. Due to the principle of duality, dual problems, e.g. line-convex polygon, line clipping, can be solved in a similarly.

## INTRODUCTION

In many applications Point-in-Polygon and Point-in-Polyhedron tests are used not only for detection or point localization, but also to speed up algorithms, in general. If many points are to be tested and the given polygon or polyhedron are constant, a preprocessing can be used to increase efficiency of the test. In the non-convex case, i.e. polygon or polyhedron are non-convex, the computational complexity is $O(N)$ in both cases. If a convex polygon in $E^2$ is given, then algorithms with $O(N)$ or $O(\log N)$ complexities can be used. If a convex polyhedron in the $E^3$ case is given and information on neighbor faces is available then the algorithm with $O_{expected}(\sqrt{N})$ expected complexity [11] can be used, otherwise an algorithm with $O(N)$ must be used instead. If number of faces $N$ is higher, computational time might not be acceptable for higher number of tested points.

Notation used:
- $N$ – number of edges of a polygon or number of faces of a polyhedron
- $M$ – number of tested points
- $x_T$ – point inside of a convex polygon or convex polyhedron, e.g. a center of gravity estimation
- $x_i$ – $i^{th}$ vertex of the given convex polygon
- $f_i$ – $i^{th}$ edge of the given convex polygon, resp. $i^{th}$ face of the given polyhedron
- AABB – Axis Aligned Bounding Box

## POINT IN CONVEX POLYGON AND POLYHEDRON TEST

The simplest algorithm for Point-in-Convex Polygon test is of $O(N)$ complexity. It is expected that edges are oriented clockwise or anticlockwise and edges $f_i$ define half-planes determined as $F_i(x) \geq 0$ so that

$$F_i(x_T) = a_i x_T + b_i y_T + c_i \geq 0 \qquad \forall i = 0, \dots, n-1 \qquad (1)$$

If $F_i(x) \geq 0 \; \forall i = 0, \dots, n-1$ then the tested point $x$ is inside of the convex polygon, Fig.1. This algorithm is of $O(N)$ computational complexity. In the case of Point-in-Convex Polyhedron test, the algorithm is similar as faces of the given polyhedron are used only instead of edges, i.e.

$$F_i(x_T) = a_i x_T + b_i y_T + c_i z_T + d_i \geq 0 \qquad \forall i = 0, \dots, n-1 \qquad (2)$$

where $i$ means the $i^{th}$ face of the given polyhedron. If $F_i(x) \geq 0$ for $\forall i = 0, \dots, n-1$, then the tested point $x$ is inside of the convex polyhedron.

Due to convexity of the given polygon and known order of vertices, which is *substantial*, an algorithm with $O(\log N)$ computational complexity exists [14] as bisection over indexes can be used.

Let us select the point $x_0$ as a reference point and define half-spaces $G_i(x) \geq 0$, Fig.2.

$$G_i(x) = a_i x + b_i y + c_i \geq 0 \qquad\qquad i = 0, \ldots, n-1 \qquad (3)$$

where: $i$ means a half-space given by the vertices $x_0$ and $x_i$.

If the point $x$ inside of the wedge given by the vertices $x_{n-1} x_0 x_1$ a new vertex $x_k = x_{\lfloor n/2 \rfloor}$ is taken. Then the original wedge $x_{n-1} x_0 x_1$ is split to two wedges $x_{n-1} x_0 x_k$ and $x_k x_0 x_1$ and the given point $x$ is tested against $G_i(x)$ halfspace. For the next step is taken the wedge in which the point $x$ lies. If the wedge is elementary, i.e. $x_{k+1} x_0 x_k$ and the point $x$ is inside, the final test whether the point $x$ is inside of the triangle $x_{k+1} x_0 x_k$ using $F_k(x) \geq 0$ has to be made.

It can be seen that this algorithm is of $O(\log N)$ computational complexity.

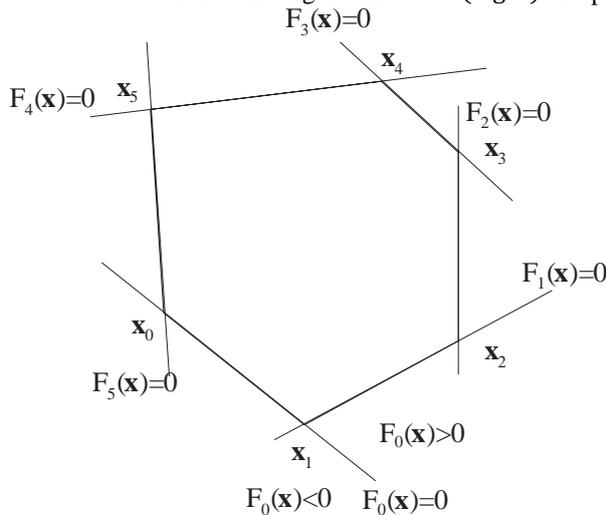
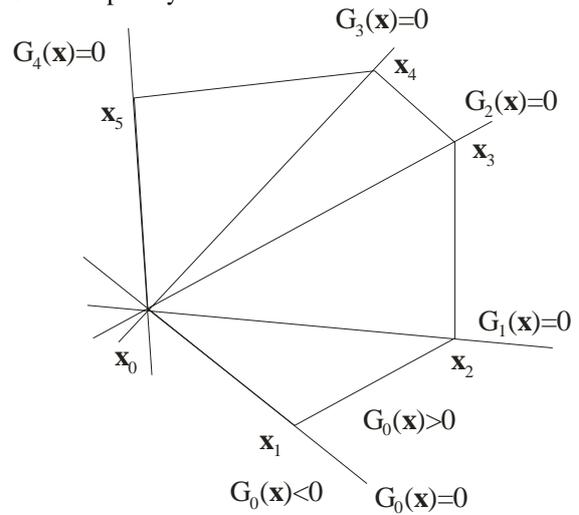

**FIGURE 1.** Point-in-Convex Polygon algorithm with $O(N)$ complexity

**FIGURE 2.** Point-in-Convex Polygon algorithm with $O(\lg N)$ complexity

It should be noted that in the $E^3$ case, *there is no ordering of vertices or faces defined*, i.e. in the case of convex polyhedron, and therefore a similar algorithm for Point-in-Convex Polyhedron test *does not exist*. If information on neighbor faces is available, an algorithm with $O_{expected}(\sqrt{N})$ expected complexity can be used [11].

Another approach [7] is based on an analogy with the scan-line algorithm. In this case the given polygon is preprocessed and then represented by slabs, Fig.3, and for each slab a list of active edges is kept. The preprocessing is of $O(N \lg N)$ complexity due to $y$-coordinate sorting. In the runtime a relevant slab has to be found, which is of $O(\lg N)$ complexity due to binary search of $y$-coordinate.

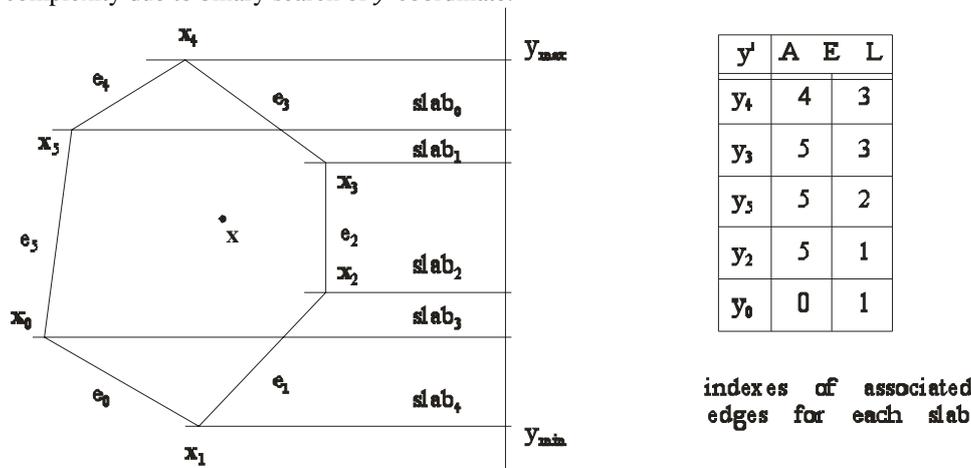

**FIGURE 3.** Point-in-Convex Polygon algorithm with $O(N \lg N)$ preprocessing and $O(\lg N)$ run-time complexity

# POINT-IN-CONVEX POLYGON WITH $O(1)$ COMPLEXITY

The algorithm [7] is on $O(\lg N)$ run-time complexity, however if all slabs are "infinitely slim" then the run-time can be decreased to $O(1)$ run-time complexity, Fig.4, as the index $i$ of a slab can be determined as

$$i = \left\lfloor \frac{y - y_{min}}{y_{max} - y_{min}} N_{slabs} \right\rfloor \quad (4)$$

where: $N_{slabs}$ is a number of slabs depending on the minimal polygon edge length. Now, for each slab a list of active edges is kept. Of course, the preprocessing complexity is becoming quite high [12].

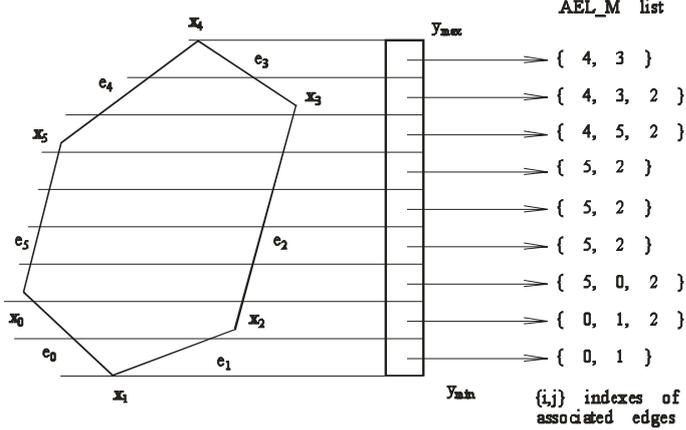 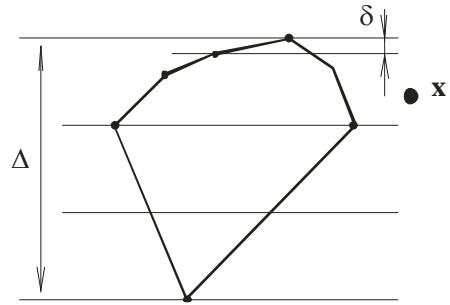

**FIGURE 4.** Point-in-Convex Polygon algorithm with $O(1)$ complexity     **FIGURE 5.** Dependence of the slab thickness

However, the thickness of all slabs is not infinitely small. The thickness of all slabs is determined by geometric properties of the given convex polygon, actually by the smallest distance in the $y$ coordinates, see Fig.5.

# NEW PROPOSED ALGORITHM

The proposed algorithm is simple as it is based on combination of the scan-line algorithm, "polar" coordinate system and mapping to the AABB. The first step is an estimation of the centroid of the given convex polygon or its estimation. For the point tested, a relevant octant and relevant positional slab $e_i$ is determined. For each slab $e_i$ index of the polygon edge intersected by a ray $x_T x$ is retrieved. If the point is on the left hand side of the edge $f_i$, then the point $x$ is inside of the given convex polygon. It can be seen that the minimal length of a slab $e_i$ is determined by the shortest polygon edge length. The slabs $e_i$ sizes might be different, e.g. the slab index might be computed as $i = \lfloor y/x \rfloor = q * \tan \varphi$, where $q$ is a multiplicative constant and as the $\tan \varphi$ function is monotonic on $\langle 0, \pi/4 \rangle$. The virtual bounding box can be set as a bounding box of the given convex polygon. It should be noted that there might be up to 2 edges associated with a slab if the slab contains a polygon vertex.

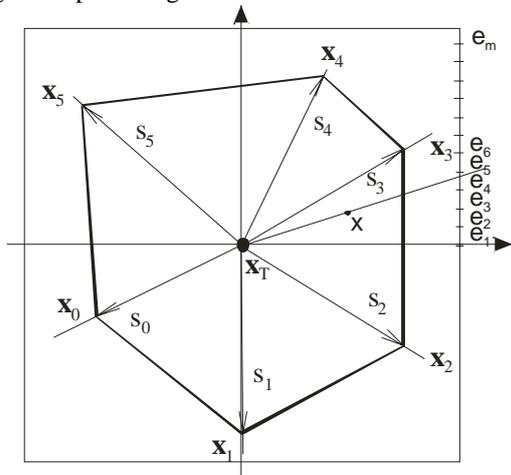 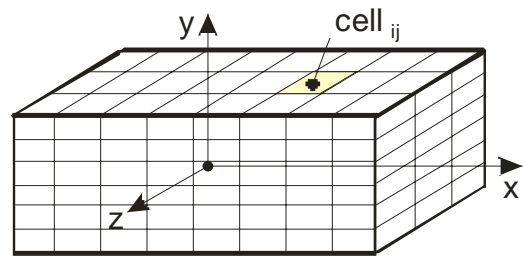

**FIGURE 6.** Polar slabs decomposition of a convex polygon     **FIGURE 7.** "Spherical" decomposition of a convex polyhedron (actually mapping to a bounding cube is made)

# POINT-IN-CONVEX POLYHEDRA WITH $O(1)$ COMPLEXITY

The above presented algorithm can be easily modified for the $E^3$ case, i.e. for the case of Point-in-Convex Polyhedron test. The convex polyhedron is bounded by a virtual three dimensional bounding box. A rectangular space subdivision is "spherically" applied on each face of the bounding box resulting to orthogonal planar cells. For each cell a list of associated faces of the convex polyhedron is kept. The cell size is related to the smallest face of the convex polygon. In the vast majority the list contain only one face index, except of cases when a vertex or edge of the given convex polygon "fall" into the cell.

## CONCLUSION

This paper presents a new approach for Point-in-Convex Polygon in E² and Point-in-Convex Polyhedron in E³ tests based on polar and "spherical" space subdivisions using AABB. In the case of convex polygon, resp. convex polyhedron, the processing time is $O(1)$. The pre-processing time depends on geometrical properties of the given polygon, resp. polyhedron. In the case of a convex polygon, the preprocessing time is related to the minimal edge length. In the case of a convex polyhedron, preprocessing time is related to the smallest area of the polygon face.

The proposed algorithms are to be used if many points are to be processed and the given polygon, resp. polyhedron is constant. The proposed algorithms were experimentally verified and experimental results proved expected preprocessing and run-time complexity. It should be noted that actual timing is sensitive to hardware used and data caching due to extreme simplicity of the run-time part.

## ACKNOWLEDGMENT

The author thanks to students and colleagues at the University of West Bohemia for recommendations, constructive discussions, and hints that helped to finish the work. Many thanks belong to the anonymous reviewers for their valuable comments and suggestions that improved this paper significantly. This research was supported by the Ministry of Education of the Czech Republic – project No.LH12181.